\documentclass{aa}
\usepackage{graphicx}
\usepackage{natbib}
\usepackage{aas_macros}

\begin{document}

\title{The K-band spectrum of the Cataclysmic Variable RXJ~0502.8+1624 (Tau~4)}

\author{S.P. Littlefair\inst{1} \and V.S. Dhillon\inst{1}
\and E.L. Mart\'{i}n\inst{2}}

%\offprints{S.P. Littlefair}

\institute{Department of Physics and Astronomy, University of
Sheffield, Sheffield S3 7RH, UK\\
\email{S.Littlefair@sheffield.ac.uk}
\and
Instituto de Astrof\'{i}sica de Canarias, 38200 La Laguna, Tenerife, Spain\\
University of Central Florida, Department of Physics, PO Box 162385, Orlando,
FL 32816-2385, USA\\
}

\date{Submitted for publication \today} 

\abstract{ We present the K-band spectrum of the cataclysmic variable
RXJ~0502.8+1624 (Tau~4). The spectrum shows a broad, smooth hump, with
no absorption lines from the secondary star visible. This result
indicates that the infrared light of this system is dominated by
cyclotron emission, and, in combination with the optical spectrum and
X-ray properties, suggests that Tau 4 is a polar-type cataclysmic
variable (CV).

The system was chosen for study because the broadband JHK colours of
Tau 4 are consistent with an L-type dwarf, suggesting that this
system might harbour an elusive sub-stellar secondary star. The result
presented here, along with the recent discovery of cyclotron emission
in the cataclysmic variable EF Eri, suggests that care must be taken when
using the broadband JHK colours of CVs when targeting searches for
sub-stellar secondary stars. 
}

\maketitle

\section{Introduction}
\label{sec:introduction}

Cataclysmic variables (CVs) are semi-detached binary stars consisting
of a white dwarf primary and a Roche-lobe filling secondary star.
Evolutionary models predict that as the secondary transfers mass to
the white dwarf, the period of the binary star decreases. Eventually,
the mass of the secondary drops below the hydrogen-burning limit and
the secondary star becomes degenerate. This change in the structure of
the secondary star means that further mass loss is accompanied by an
increase in the orbital period \cite[see][for example]{kolb93}. This
is often used to explain the orbital period minimum which is observed
in CVs at around 80 minutes. Evolutionary models \cite[e.g]{howell97}
suggest that around 70 per cent of CVs should have passed the orbital period
minimum and should therefore contain sub-stellar secondary stars.

The observational evidence for these sub-stellar secondaries was
critically examined by ourselves in an earlier paper
\citep{littlefair03}. We concluded that despite their predicted
abundance amongst the CV population there is as yet no direct evidence
that any CVs possess a sub-stellar secondary star\footnote{Indirect
evidence exists from modelling of the SED, use of a superhump
period-mass ratio relationship and radial velocity studies of the
emission lines}. In that paper, we remarked that there were a number
of CVs whose JHK colours were highly suggestive of a sub-stellar
secondary, and suggested that these CVs be followed up with infrared
spectroscopy. One such system is Tau~4.

RXJ~0502.8+1624 (Tau~4) was first identified as a CV by
\cite{motch96}, who observed selected sources from the ROSAT all sky
survey (RASS). Their optical spectrum showed narrow Balmer
and Helium emission lines, including the HeII $\lambda$4686 line. The
emission lines showed a broad component underlying a strong, narrow
peak. The X-ray hardness ratios and the ratio of HeII to H$\beta$
equivalent widths \citep{motch96}, is typical of the polar class of
magnetic CV, as is the characteristic emission line structure
\citep{warner95a}. These properties, taken together, are good
indicators that Tau 4 is a polar. The infrared counterpart of Tau~4
was securely identified by \cite{hoard02}. They found infrared colours
of $J-H=0.95$ and $H-K=0.88$, and a K-band magnitude of $K=14.66$,
consistent with those of an L3 dwarf star (see figure~\ref{fig:cmd}),
suggesting that a sub-stellar secondary may dominate the infrared
light of this system.  Here we present the K-band spectrum of
Tau~4. The observations and data reduction techniques applied are
outlined in section~\ref{sec:obs}. The spectrum is presented in
section~\ref{sec:results} and the results are discussed in
section~\ref{sec:discussion}. Our conclusions are presented in
section~\ref{sec:conclusions}.

\section{Observations and Data Reduction}
\label{sec:obs}
On the night of 2002 November 25 we obtained 2.0560--2.4730~$\mu$m
($\sim270$~km\,s$^{-1}$ resolution) spectra of the CV Tau~4 with
NIRSPEC \citep{mclean98} on the 10-m Keck-II telescope on Mauna Kea,
Hawaii. A total exposure time of 900 secs was obtained between
airmasses of 1.02--1.04 in photometric conditions. The seeing was
approximately 0.7 arcseconds and the slit width was set to 0.76
arcseconds. Observations of the A2V star HD~2127 were also taken to
correct for the effects of telluric absorption and to provide flux
calibration. Both stars were observed with a slit position angle of
zero degrees. Using the equations of \cite{fillipenko82} to estimate
the effects of differential refraction we estimate the slit losses at
2.4$\mu$m relative to 2.2$\mu$m to be less than 1\%. We therefore
conclude that wavelength-dependent slit losses are not significant for
our data.

NIRSPEC introduces curvature and distortion in both the spatial and
dispersion directions. Prior to extraction of spectra, these effects
were removed using the {\sc wmkonspec} package in {\sc iraf}.
Following the removal of the distortion, the nodded frames were
subtracted, the residual sky removed by subtracting a polynomial fit,
and the spectra extracted. There were two stages to the calibration of
the extracted spectra. The first was the calibration of the wavelength
scale using argon arc-lamp exposures; the fifth-order polynomial fits
to the arc lines yielded an error of less than 0.13\AA\ (rms).  The
second step was the removal of telluric features and flux
calibration. This was performed by dividing the spectra to be
calibrated by the spectrum of the A2V standard, with its prominent
stellar features interpolated across. We then multiplied the result by
the known flux of the standard at each wavelength, determined using a
black body function set to the same effective temperature and flux as
the standard. As well as providing flux calibrated spectra, this
procedure also removed telluric absorption features from the object
spectra. A final, average spectrum was produced by co-adding the
spectra in the rest frame of the binary centre-of-mass. The resultant
spectrum resulted in a signal to noise ratio which varies smoothly
from 35 around 2.1$\mu$m to 20 at 2.35$\mu$m.

\section{Results}
\label{sec:results}
\begin{figure}
  \includegraphics[scale=0.5]{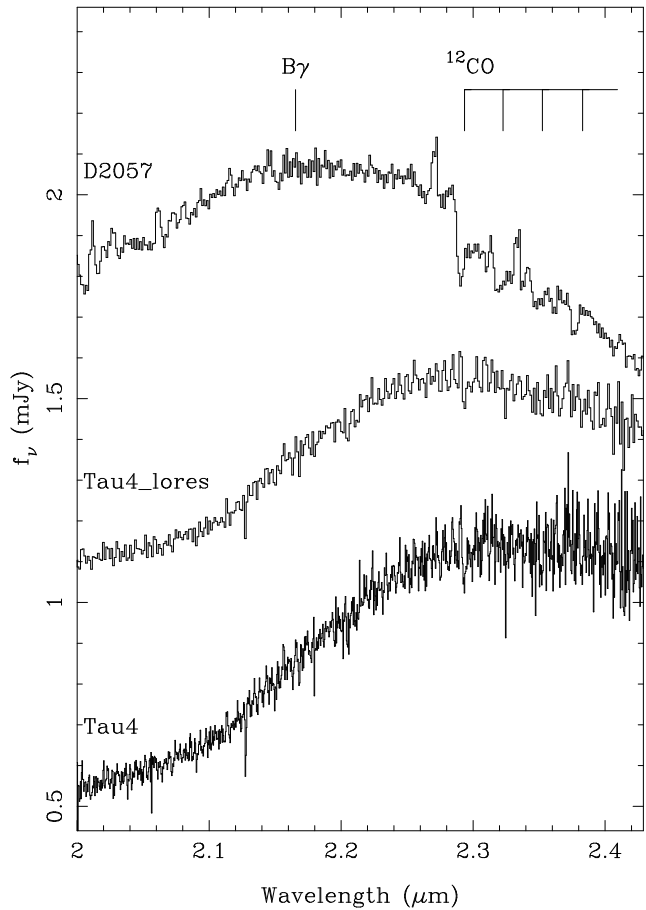}
\caption{K-band spectrum of Tau 4 (bottom). Also shown is the
\cite{kendall04} spectrum of the L1.5 dwarf DENIS-P 205754.10-025229.9
(top), and our spectrum of Tau 4, binned to match the resolution of
the brown-dwarf spectrum (middle). The spectra have been normalised,
and an offset added for clarity.}
\label{fig:spectrum}
\end{figure}

Figure~\ref{fig:spectrum} shows the Keck spectrum of Tau~4 in the
K-band. The spectrum is dominated by a smooth hump-shaped
continuum. There is no evidence for the emission line of
Brackett-$\gamma$.  The spectrum shows possible evidence for $^{12}$CO
absorption lines from the secondary star. The reality of these
features is highly uncertain: the spectrum is quite noisy and the
$^{12}$CO lines occupy a region of the spectrum which can be badly
affected by telluric absorption.  In addition, the continuum shape of
the spectrum is not consistent with the continuum shape of late-M or
L-type dwarfs \cite[see][for example]{mclean03}. The continuum shape
is, however, consistent with cyclotron emission
\cite[see][]{ferrario93}. Circularly polarised cyclotron emission is
routinely used to measure the field strengths in polars, where the
magnetic field is typically 10--80\,MG. By contrast, only three
intermediate polars (which show lower field strengths) have to date
shown significant polarisation, and no intermediate polar has had its
field strength measured from cyclotron harmonics
\citep{buckley00}. Detection of cyclotron emission in Tau~4 is
therefore strong confirmation of the polar nature of this object.
 
\section{Discussion}
\label{sec:discussion}
\begin{figure}
  \includegraphics[scale=0.35,angle=-90]{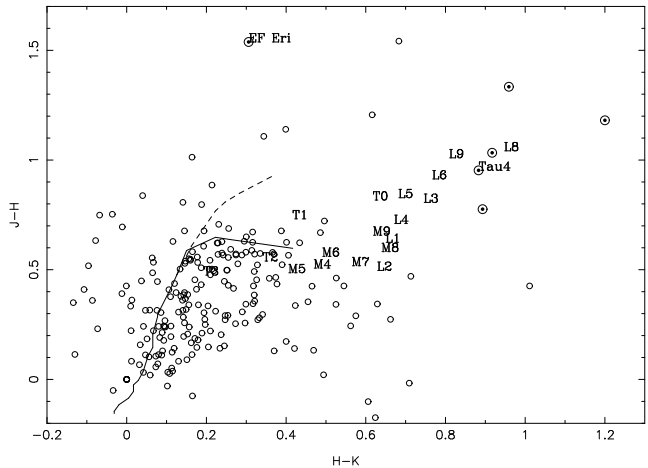}
\caption{Infrared colour-colour diagram for CVs and late-type
dwarfs. The open dots show the CVs from the 2nd incremental data release of
the 2MASS survey \citep{hoard02}. Those CVs whose colours suggest they
might possess sub-stellar secondary stars are marked with large dots
surrounded by circles. The positions of late-type dwarfs (from
\citealt{leggett02}) in the colour-colour diagram are represented by a
text string indicating their spectral type.  The solid curve shows the
position of the main sequence from spectral-types O9 to M5; the dashed
curve shows the position of the giant sequence from spectral-types G8
to M5 \citep{allens00}. The colours of the main sequence and late-type
stars were put on the 2MASS photometric system using the
transformations of \citet{carpenter01}.}
\label{fig:cmd}
\end{figure}
Figure~\ref{fig:cmd} shows the colours of CVs from the 2nd incremental
data release of the 2MASS survey \citep{hoard02}. One thing that is
immediately apparent from this diagram is that there are a small
number of CVs whose infrared colours are consistent with those of
late-type dwarf stars. There are also several stars, such as EF
Eri, whose infrared colours are very different from the population as
a whole. \cite{littlefair03} suggested that these systems were good
candidates for CVs with sub-stellar secondary stars. Tau 4 is one of
these objects.

From the K-band spectrum in figure~\ref{fig:spectrum}, however, it is
apparent that the infrared light in Tau~4 is dominated by cyclotron
emission. This result shows that the characteristic humps of cyclotron
emission can mimic the near-IR colours of a late-type secondary star,
and must inevitably cast doubt on the nature of the other CVs in this
region of the colour-colour diagram. This result is further
strengthened by the evidence presented in \cite{harrison04}. The
authors present time-resolved H and K-band spectroscopy of EF Eri,
whose infrared colours are also highly unusual (see
figure~\ref{fig:cmd}). These spectra show that the infrared light in
EF Eri is also dominated by cyclotron emission. Although
\cite{harrison04} may have detected faint traces of the secondary star
around phase 0, the secondary star certainly does not make a {\em
significant} contribution to the near-IR light in this system. The
conclusion is that EF Eri's unusual near-IR colours are not due to the
secondary star, but to cyclotron emission.

Tau~4 and EF Eri represent the only two systems with unusual IR
colours in figure~\ref{fig:cmd} for which infrared spectra have been
obtained. Given that the infrared light in both these systems is
dominated by cyclotron emission it is now clear that not all CVs which
occupy the same region of the $J-H$ vs $H-K$ colour-colour diagram as
late-type stars possess late-type secondaries which dominate the
near-IR light. This will inevitably reduce the efficiency of using
near-IR colours as a selection method for CVs with late-type
secondaries, but does not mean there are no CVs in which a late-type
secondary dominates the near-IR light.

\section{Conclusions}
\label{sec:conclusions}
We present the K-band spectrum of the CV Tau~4. The
spectrum shows a broad, smooth continuum hump, which is most likely to
be  cyclotron emission. The K-band spectrum, in combination
with the emission lines seen in the optical spectrum, suggest that
Tau~4 belongs to the polar class of magnetic CVs. Given the
detection of cyclotron emission in Tau~4 and EF Eri, we conclude that
near-IR colours alone are not sufficient to infer the presence of
a late-type secondary star.

\section*{\sc Acknowledgements}
SPL is supported by PPARC. The authors acknowledge the data analysis
facilities at Sheffield provided by the Starlink Project which is run
by CCLRC on behalf of PPARC. This material is based upon work
supported by the National Science Foundation under grant 0205862. Any
opinions, findings, and conclusions or recommendations expressed in
this material are those of the author(s) and do not necessarily
reflect the views of the National Science Foundation. The authors wish
to extend special thanks to those of Hawaiian ancestry on whose sacred
mountain of Mauna Kea we are privileged to be guests. Without their
generous hospitality, the Keck II telescope observations presented
therein would not have been possible.

\bibliographystyle{aa} \bibliography{abbrev,refs}

\end{document}